\documentclass[prl,aps, english, twocolumn, hyperref]{revtex4}

\usepackage{graphicx}

\usepackage{amssymb, amsmath, color, rotating}
\begin{document}

\title{Long time non-equilibrium dynamics of binary Bose condensates}

\author{Stefan S. Natu}

\email{ssn8@cornell.edu}

\affiliation{Condensed Matter Theory Center and Joint Quantum Institute, Department of Physics, University of Maryland, College Park, Maryland 20742-4111 USA}

\author{S. Das Sarma}

\affiliation{Condensed Matter Theory Center and Joint Quantum Institute, Department of Physics, University of Maryland, College Park, Maryland 20742-4111 USA}

\begin{abstract}
We explore the out-of-equilibrium temporal dynamics of demixing and phase separation in a two-dimensional binary Bose fluid at zero temperature, following a sudden quench across the miscible-immiscible phase boundary. On short timescales, the system rapidly settles into a steady state characterized by short-range correlations in the relative density. The subsequent dynamics is extremely slow: domains of the relative density appear to grow with time, however the rate of growth is much slower than that predicted by conventional theories of phase ordering kinetics. Moreover, we find that the growth dynamics slows down with increasing time, and is consistent with logarithmic growth.  Our study sheds light on ongoing investigations of how isolated quantum systems approach equilibrium, and indicates that studying the quantum phase diagram of the binary Bose fluids following a quench, may be difficult due to equilibration problems.
\end{abstract}

\maketitle

The question of the growth of an order parameter following a sudden quench from a disordered to an ordered phase is a fundamental one \cite{brayarticle, hhalperin}. The picture that has emerged is one of domain nucleation and coarsening. For example, in a binary system such as a binary alloy, phase ordering kinetics proceeds via spinodal decomposition, whereby the short and long time dynamics can be attributed to fast and slow modes in the system. On short timescales, the system develops domains whose characteristic size is determined by the fastest growth modes in the system. On intermediate times, these domains merge and grow.  At long times, one arrives at a coarse-grained, low energy description that is independent of the microscopic details of the system, and is controlled only by the applicable conservation and symmetry principles. Dynamical scaling properties of various physical quantities under non-equilibrium conditions  comprise a vast interdisciplinary subject in classical statistical mechanics \cite{brayarticle, hhalperin}. 

With the recent developments in the field of ultra-cold gases, there has been an explosion in interest, in asking analogous questions for non-equilibrium dynamics across a \textit{quantum} phase transition \cite{polkovnikovreview}. Unlike their solid state counterparts, ultra-cold gases are to an excellent approximation, isolated from their environment, which opens up the possibility of observing quantum coherent dynamics on long timescales \cite{greiner}.  Furthermore the ability to probe these gases locally \cite{greinerimg, blochimg} and globally \cite{andrews, chengimg} using imaging techniques may offer new insights into phase ordering kinetics. Here we study the dynamics of domain growth in a binary Bose condensate at zero temperature, suddenly quenched across the miscible-immiscible phase boundary \cite{cornellbinary, ketterlebinary, ao2, spielmanbinary}. Studying such non-equilibrium quantum dynamics in solid state systems is essentially impossible because of environmental decoherence and the fast timescales (often femtoseconds) involved in dynamics. 

Conventional theories of phase ordering kinetics argue that at long times ($t$), all the information about the system is contained in a single observable: the size of the order parameter domains $(L(t))$. Understanding how this length scale grows with time yields insight into the low energy physics of the system, and the underlying conservation laws. In most cases, the length scale grows as a power-law $L(t) \sim t^{1/z}$, where $z$ is the dynamical scaling exponent and is the quantity of interest \cite{brayarticle}. 
For example, diffusive dynamics of a scalar order parameter yields a growth law $L(t) \sim t^{1/2}$ (Model A) \cite{brayarticle, hhalperin},  whereas if the order parameter is globally conserved, $L(t) \sim t^{1/3}$ (Model B)\cite{lifshitz, brayprl, voorhees, huse}. In fluids where advection dominates particle transport, Siggia argued that $L(t) \sim t$ \cite{siggia}. Depending on the underlying conservation laws, other possible growth laws could apply in different situations. 

Here we consider a binary Bose condensate where the total density of each of the fluids is independently conserved. Furthermore, unitary evolution also conserves the total energy. We simulate the dynamics of phase separation using a Gross-Pitaevskii equation, which captures all the relevant physics of weakly interacting condensates at zero temperature. We find that domains of each fluid grow much more slowly than the above theories predict (\textit{i.e.} dynamical scaling exponents of $z = 1, 2$ or $3$). On the timescales we simulate, the dynamics is consistent with an asymptotic logarithmic growth. Similar anomalous growth laws are known to occur in systems exhibiting glassy dynamics (See Ref.~\cite{evans} and references therein).

Much of the theoretical work on the long time dynamics following a quantum quench has been limited to one-dimensional systems where powerful analytical and numerical techniques exist \cite{essler, rigol1, kolodrubetz, kollath, roux, yukalov}. In one dimension, the absence of true long-range order, combined with phase-space constraints on kinematics, can lead to long relaxation times, or relaxation to athermal steady states with short range correlations \cite{kinoshita, rigol2, rigol3, cardy}. By contrast, in higher dimensions the question of whether quasi/true long range order is established after a quench (either quantum or thermal) becomes relevant \cite{sauquench, natubogquench, barnett, mukherjeemoore, damle, lamacraft}. In the two-dimensional system we consider, we find that the two fluids \textit{never} completely phase separate, and only local correlations in the relative density are appreciable at long times. 

Our system consists of a homogeneous, binary Bose mixture (denoted by $a$ and $b$) of equal mass in two dimensions interacting with a contact interaction:
\begin{equation}\label{ham}
{\cal{H}} = -\sum_{i}\Psi^{\dagger}_{i}\Bigl(\frac{\hbar^{2}\nabla^{2}}{2m} -\mu_{i}\Bigr)\Psi_{i} + \sum_{i, j}\frac{g_{ij}}{2} \int d^{2}\textbf{r} \Psi^{\dagger}_{i}\Psi^{\dagger}_{j}\Psi_{j}\Psi_{i}
\end{equation}
where the indices $i, j$ run over $\{a, b\}$. Here $\Psi_{i}$ denotes the annihilation operator for the bosonic species $i \in \{a, b\}$, and $\mu_{i}$ is the chemical potential which fixes the density of each species. The interaction coefficients $g_{aa}, g_{bb}$ and $g_{ab}$ are all assumed to be positive, and for simplicity, we set $g_{aa} = g_{bb} = g$ and $\mu_{a} = \mu_{b} = \mu$.

At zero temperature, the atomic wave-function of each species can be expressed in terms of a c-number $\psi_{i}$ which obeys the familiar Gross-Pitaevskii equation (GPE) \cite{pethick}. For the coupled system this takes the form:
\begin{equation}\label{gpe}
i\partial_{t}\psi_{i} = \Bigl(-\frac{\hbar^{2}\nabla^{2}}{2m} -\mu_{i} + g_{ii}n_{i} + g_{ij}n_{j} \Bigr)\psi_{i}
\end{equation}
where the index $i \in \{a, b\}$ and $j\neq i$, and $n_{i}(\textbf{r}) = |\psi_{i}(\textbf{r})|^{2}$ is the density of each species. 

The mean-field physics of these equations is well understood in homogeneous and trapped systems \cite{hoshenoy, ao1}. As ${\cal{H}}$ contains no spin-flip terms, the total particle number of each species is independently conserved. This leads to two mean-field ground states: for $g_{ab} < \sqrt{g_{aa}g_{bb}}$, the ground state is miscible, whereas for $g_{ab} > \sqrt{g_{aa}g_{bb}}$, the ground state phase separates into two domains of either species separated by a domain wall, whose width is inversely proportional to $\sqrt{g_{ab}-\sqrt{g_{aa}g_{bb}}}$ \cite{ao2}. Experiments on binary Bose condensates typically use alkali atoms such as $^{87}$Rb or $^{23}$Na, where the quantity $g_{ab}/ \sqrt{g_{aa}g_{bb}}-1$ is roughly $\sim 0.01$ \cite{cornellbinary, ketterlebinary, spielmanbinary}. However this quantity may be tunable using Feshbach resonances \cite{chengcs, huletli, cornellrb, inouye}.

The short time dynamics of domain formation in an immiscible binary condensate is well understood \cite{cornellbinary, ketterlebinary, cornellrb, spielmanbinary, tsubota, ao2, ronen}: a spin-wave instability drives the formation of domains of the relative density with a characteristic size $L_{\text{dom}} \sim  2\pi/k$, where $0 < k < k_{\text{max}} = \sqrt{2m(g_{ab} - g)n_{0}/\hbar^{2}}$ \cite{tsubota} where $g = g_{aa} = g_{bb}$ and $n_{0}$ is the typical density of each fluid. The timescale for this instability is $t_{\text{inst}} \sim 2\pi/\omega_{k_{\text{max}}}$, where $\omega_{k} = \hbar k^{2}_{\text{max}}/2m$. This picture has been confirmed in experiments and numerical simulations \cite{spielmanbinary, tsubota, ronen, cornellrb, berkeleyexpts, ramanspinor}.

By contrast, the long time dynamics is not as well understood. Misener \textit{et al.} \cite{ketterlebinary} observed long lived, short length scale domains on times $t \sim 20$s. Similar dynamics has been reported in experiments and numerical simulations by the JILA group \cite{cornellbinary, cornellrb, ronen}. Experiments and numerical studies by De \textit{et al.} \cite{spielmanbinary} find evidence for domain growth, however on the timescales of the experiment, the physics of coarsening is complicated by particle loss, trap and finite temperature effects. 

\begin{figure}
\begin{picture}(200, 180)
\put(105, 120){\includegraphics[scale=0.32]{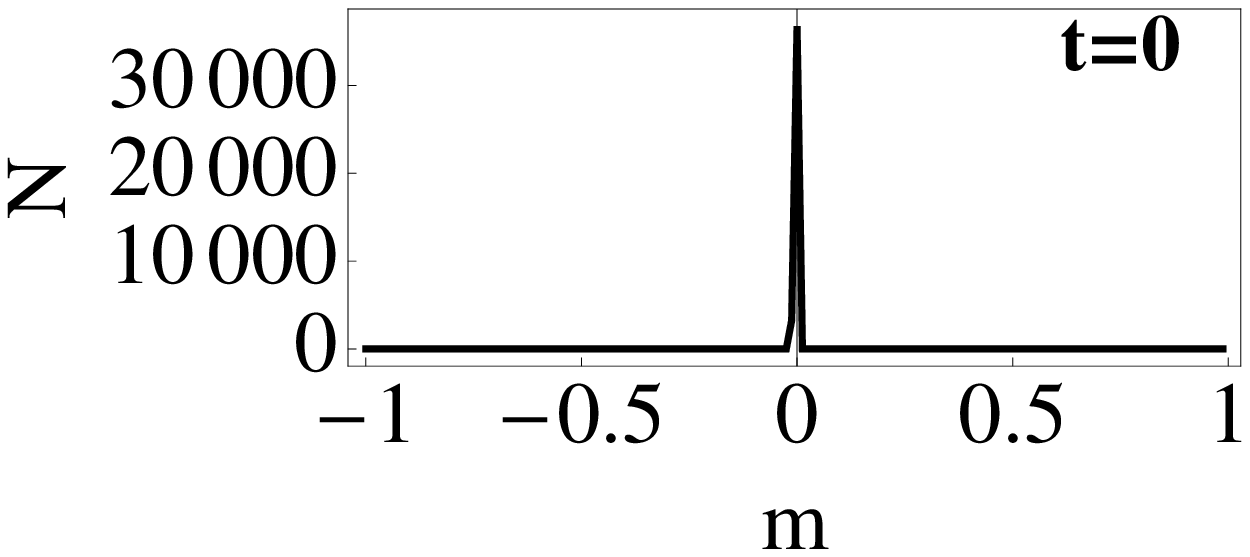}}
\put(110, 56){\includegraphics[scale=0.32]{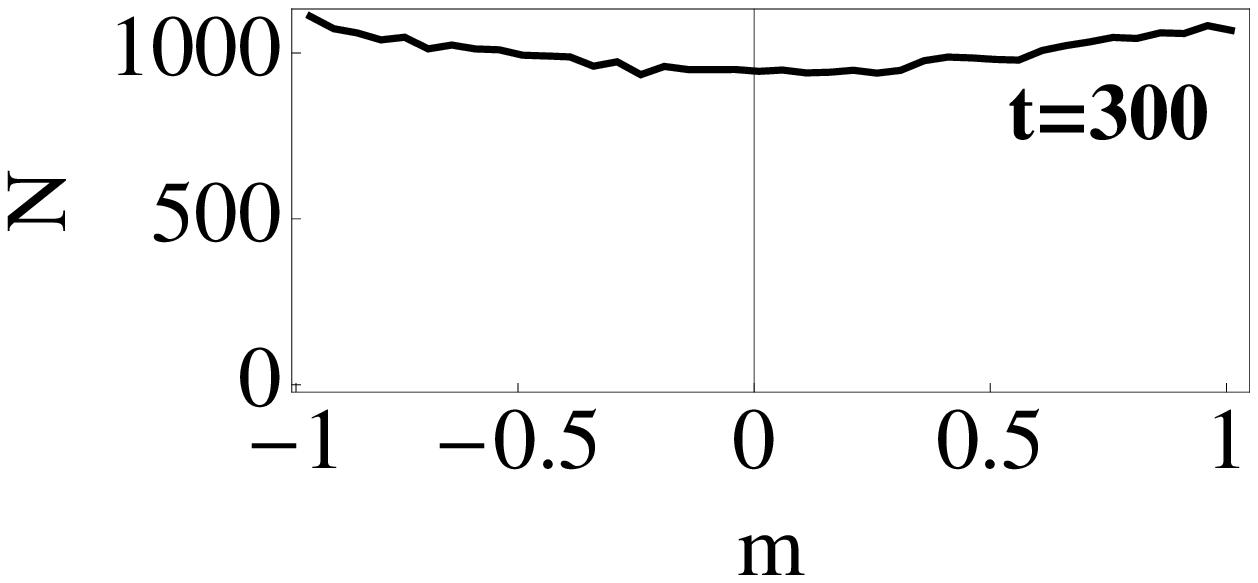}}
\put(110, -10){\includegraphics[scale=0.32]{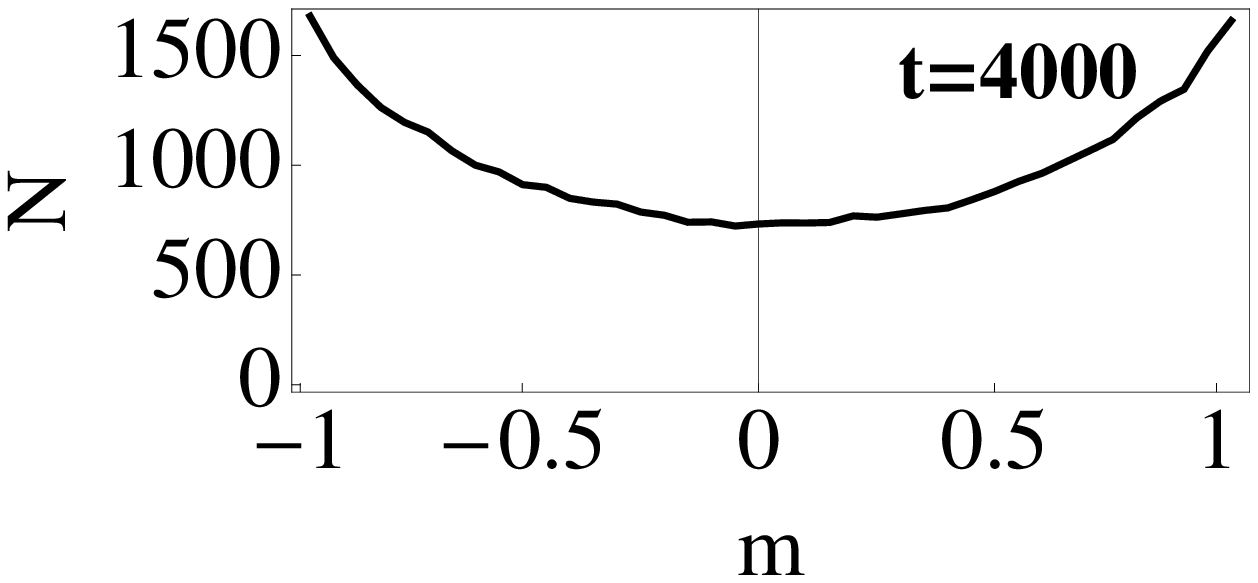}}
\put(-25, 90){\includegraphics[scale=0.35]{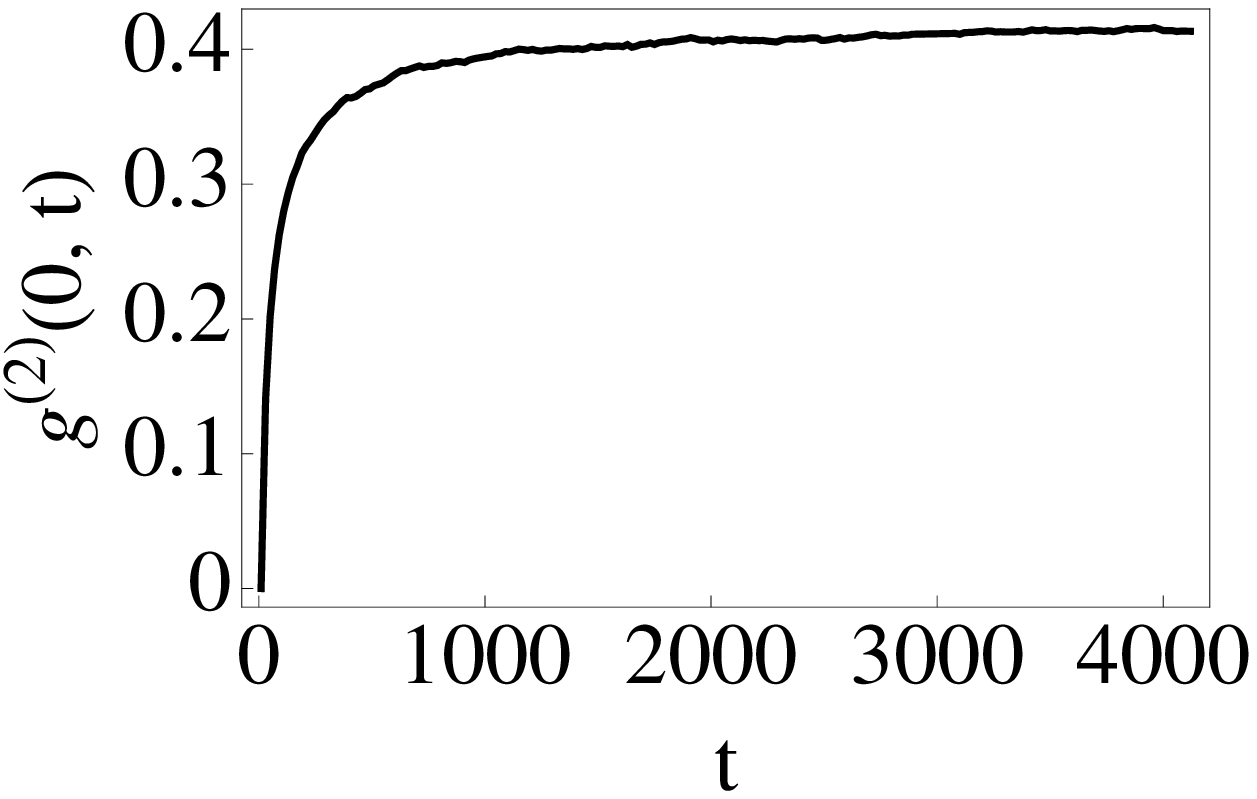}}
\put(-25, 0){\includegraphics[scale=0.35]{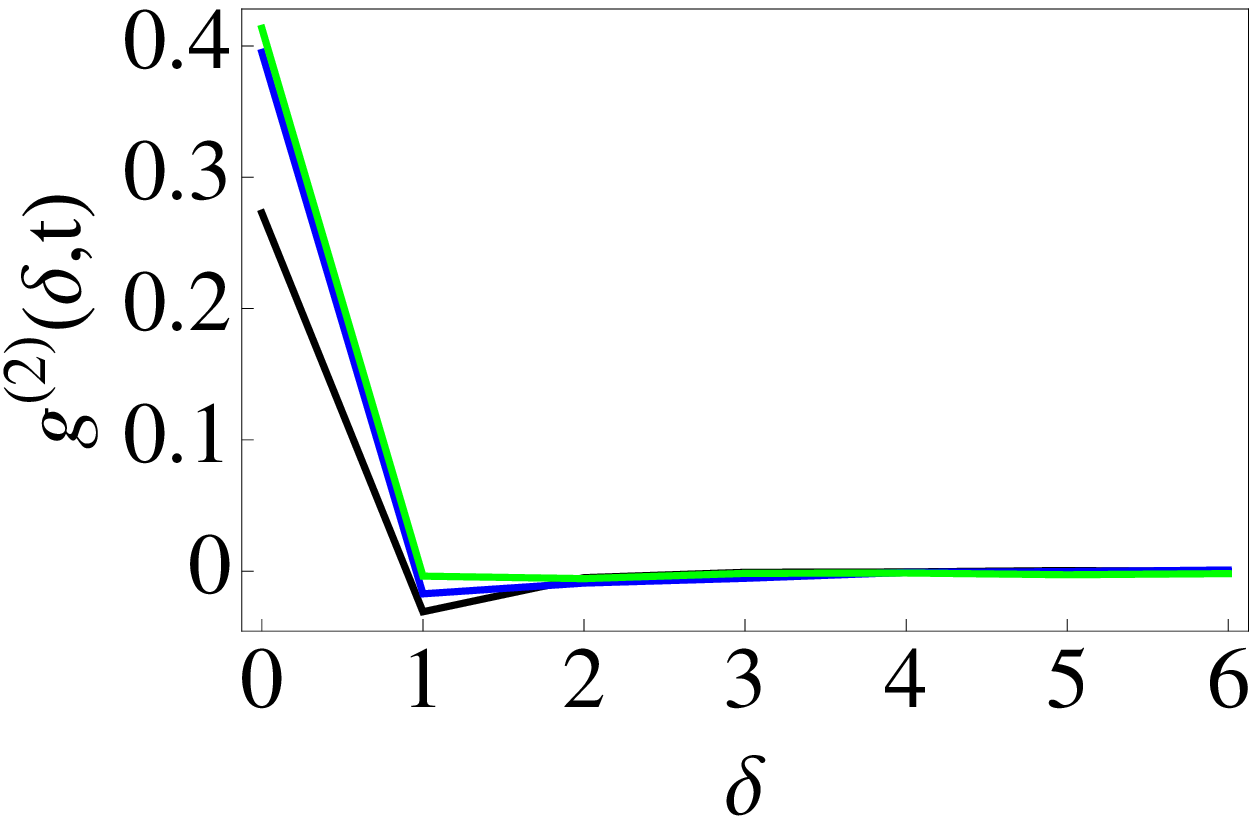}}
\end{picture}
\caption{\label{magmagcorr} (Top Left) Evolution of the local relative density-density correlation function $g^{(2)}(\delta, t) = \langle m(\textbf{r}, t)m(\textbf{r}+\delta, t) \rangle|_{\delta=0}$ where $m(\textbf{r}) =\frac{n_{a}(\textbf{r})-n_{b}(\textbf{r})}{n_{a}(\textbf{r})+n_{b}(\textbf{r})}$, plotted as a function of time for parameters $\mu = 100$, $g = 1$ and $g_{ab}-g = 0.2$.  Local correlations evolve rapidly at first and then extremely slowly on longer times. (Bottom Left) The relative density-density correlation function in space for different times $t = 300$ (black), $t = 2000$ (blue) and $t= 4000$ (green). Correlations remain short ranged even at long times. (Right Panel) Top to Bottom: Histogram showing the spread in the local polarization $m$ as a function of time. The number $N$ here is the total number of sites with polarization $m$. Following the instability, $m$ develops into a flat distribution which slowly develops broad peaks near $m= \pm 1$.}
\end{figure}

Theoretically, the story of coarsening dynamics in the Gross-Pitaevskii equation and the associated growth laws is by no means complete. The role of energy and number conserving dynamics was pointed out by Damle, Majumdar and Sachdev \cite{damle} who considered coarsening dynamics in a homogeneous Bose gas quenched from the normal phase into the Bose condensed phase. They studied the dynamics of a complex order parameter ($\langle\psi\rangle$), finding that domains with a well-defined phase grow as a power-law $L(t) \sim t^{1/z}$, with $z \sim 1$. Mukherjee, Xu and Moore \cite{mukherjeemoore} used the GPE to study coarsening dynamics at zero temperature in a spin-$1$ gas with a \textit{vector} order parameter, finding that $z \rightarrow 3$ asymptotically from above. An similar zero temperature study for the binary Bose fluid system has not been performed, and the dynamical exponent for this problem is unknown, although it is of considerable experimental interest \cite{ketterlebinary, spielmanbinary, cornellbinary}.

Below we apply the GPE to study the dynamics of a homogeneous binary Bose gas initially prepared in the miscible state. We choose $\mu = 100$ and $g = 1$ and $g_{ab} -g = 0.2$. Ignoring the kinetic energy term in Eq.~\ref{gpe}, the initial state for each species is given by $\psi_{i}(\textbf{r}) = 1/2\sqrt{\mu/(g+g_{ab})}$. As $g_{ab}>g$, the mean-field ground state is fully phase separated \cite{pethick}. We simulate the subsequent dynamics on an $L \times L$ grid where $L = 200$, and choose the time steps to be small enough to conserve the total energy and particle number in each fluid. We seed the instability by adding small random noise profile, which drives the growth of domains of the relative density. We then study the dynamics on timescales much longer than $t_{\text{inst}}$, so our results are independent of the initial conditions. 

The quantity of interest is the relative density-density correlation function defined as 
%\begin{equation}\label{magmag}
$g^{(2)}(\delta, t) = \langle m(\textbf{r}, t)m(\textbf{r}+\delta, t) \rangle$,
%\end{equation}
where the average denotes a spatial average over the sample and $m(\textbf{r}) =\frac{n_{a}(\textbf{r})-n_{b}(\textbf{r})}{n_{a}(\textbf{r})+n_{b}(\textbf{r})}$ is the local density difference (henceforth referred to as the polarization), normalized to the local density.  It  ranges from $-1 \leq m(\textbf{r}) \leq 1$. The initial state has $m(\textbf{r}) = 0$, hence $g^{(2)}(\delta, t=0) = 0$. In the fully phase separated state, we expect $g^{(2)}(\delta) \sim 1$ on length scales $0 < \delta \sim L/2$. 

In Fig.~\ref{magmagcorr}, we plot the local relative density-density correlation function $g^{(2)}(\delta = 0, t)$ as a function of time for $g_{ab}-g = 0.2$. On short timescales, a linearized analysis reveals that the correlation function grows exponentially, owing to an unstable relative density mode \cite{lamacraft, barnett}. The theory breaks down when the local density difference becomes comparable to the total density, beyond which, non-linear terms in the GPE become important. Subsequently, the correlation function grows extremely slowly with time. 

In Fig.~\ref{magmagcorr}, we plot a histogram showing the local polarization across the sample. The initial state is unpolarized, therefore the histogram is peaked at zero. Upon the onset of the instability, the histogram rapidly develops a broad distribution, where the local polarization uniformly takes on all values between $-1$ and $1$. At very long times, the local polarization develops broad peaks near $m = \pm 1$. 

We also plot $g^{(2)}(\delta, t)$ at different times for the case $g_{ab} - g = 0.2$. On short times, $g^{(2)}(\delta, t)$ is \textit{negative} at finite distances, indicating that one is more likely to find atoms of different species next to each other. This is because domains are formed by pushing unlike atoms apart. On longer times, $g^{2}(\delta)$ slowly recovers towards zero, but only local correlations are present. Similar slow dynamics was also found in numerical simulations on spin-$1$ gases by Barnett \textit{et al.} \cite{barnett}.

The data of Fig.~\ref{magmagcorr} implies that on average, the domain sizes throughout the sample are quite small. Within a domain, $g^{(2)}(\delta) > 0$, while across a domain boundary, $g^{(2)}(\delta)$ switches sign. Upon averaging over the entire sample, $g^{(2)}(\delta) \approx 0$, which means that the average domain size is only one lattice site wide. Owing to this small domain size, we cannot conclude whether or not the system is coarsening simply by studying the density-density correlation function. 

To better model the coarsening dynamics, we partition the system at time $t$ into ``domains" of size $l \times l$ where $l = \sqrt{l^{2}_{x} + l^{2}_{y}}$ whenever the absolute value of the total density difference $|m(r)|$ in a box of size $l_{x} \times l_{y}$ exceeds a cutoff $m_{\text{crit}} = 0.25$. By producing a histogram of the number of such domains as a function of $l$, we compute the average domain size at a given time. 

\begin{figure}
\begin{picture}(200, 110)
\put(-25, -10){\includegraphics[scale=0.335]{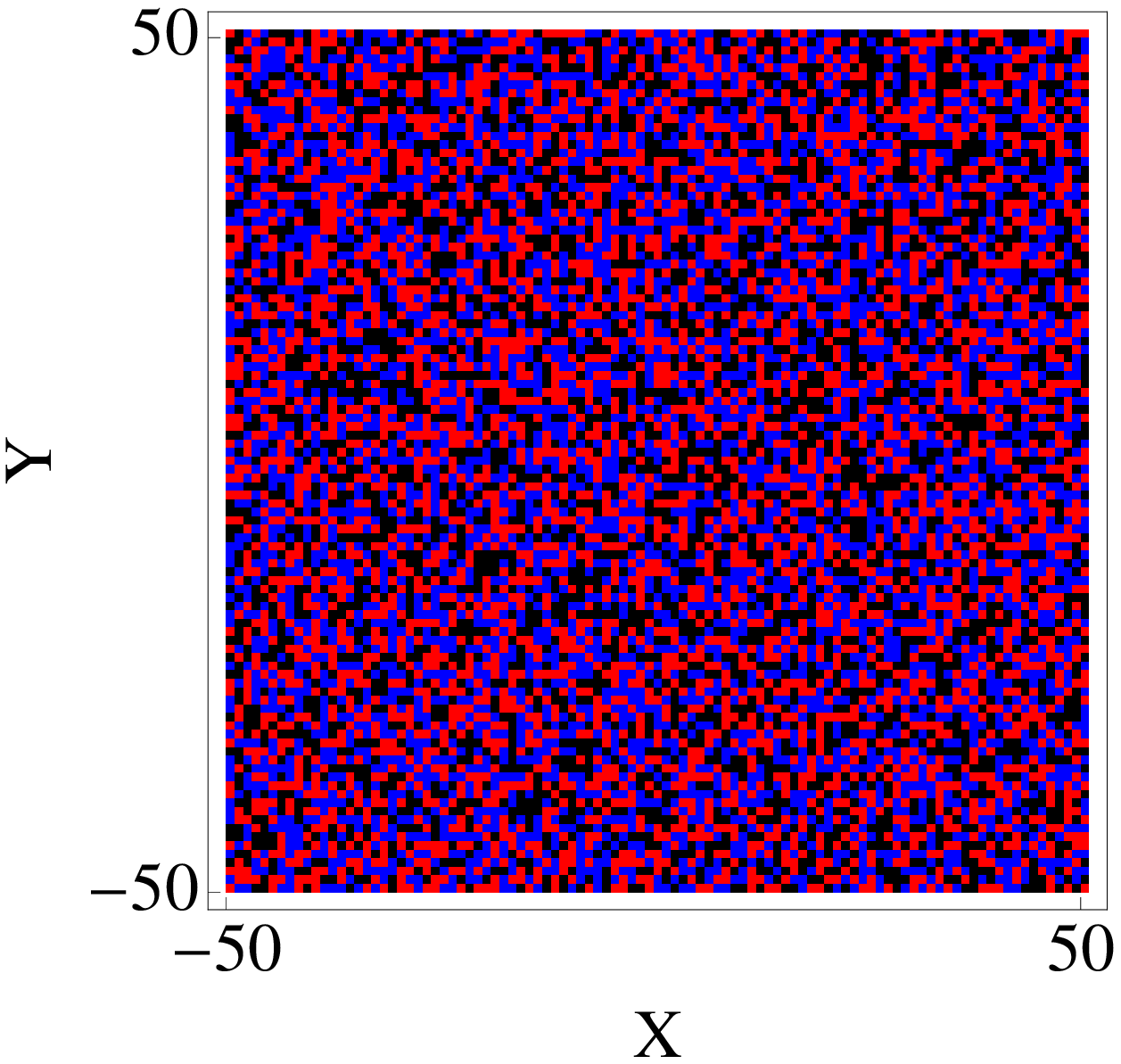}}
\put(102, -10){\includegraphics[scale=0.335]{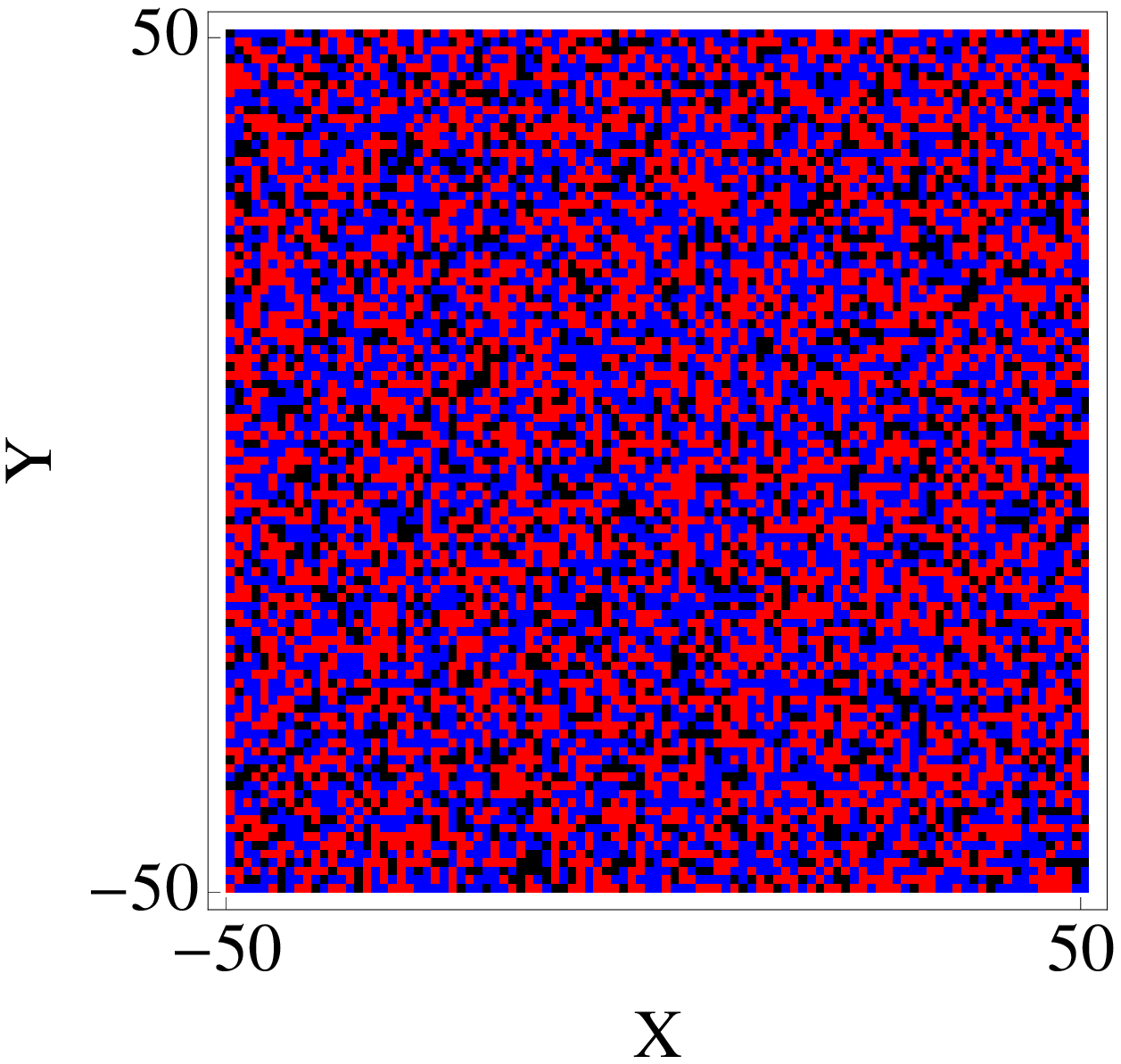}}
\end{picture}
\caption{\label{coarsening} Real space cut showing domains of the relative density at times $t = 300$ (Left) and $t= 4000$ (Right). We choose  $\mu = 100$ and $g = 1$ and $g_{ab}-g = 0.2$. The full system size is $L\times L$ where $L = 200$. The value of $m(\textbf{r}) =\frac{n_{a}(\textbf{r})-n_{b}(\textbf{r})}{n_{a}(\textbf{r})+n_{b}(\textbf{r})}$ is computed on each site and ascribed color red whenever $m(r) > 0.25$, blue whenever $m(r) < -0.25$ and black whenever $-0.25 \leq m(r) \leq 0.25$. Over time the number of black regions decreases indicating that the system is becoming more polarized locally. As is apparent from the figure, the average size of domains is growing, indicative of coarsening.}
\end{figure}

In Fig.~\ref{coarsening}, we plot the evolution of domains for different times. Here the red color indicates $m(r) > 0.25$ (domains of species $a$), while blue indicates  $m(r) < -0.25$ (or domains of species $b$). The black colors indicate values in between. For the parameters chosen in the numerical simulations, the initial spin wave instability leads to domains with a characteristic size $L_{\text{dom}} \sim 1$ lattice site. As is clear from the figure, over time, the number of black regions disappear and the red  and blue domains expand in size, indicating that the system is indeed coarsening.  

\begin{figure}
\begin{picture}(200, 80)
\put(-30, -10){\includegraphics[scale=0.35]{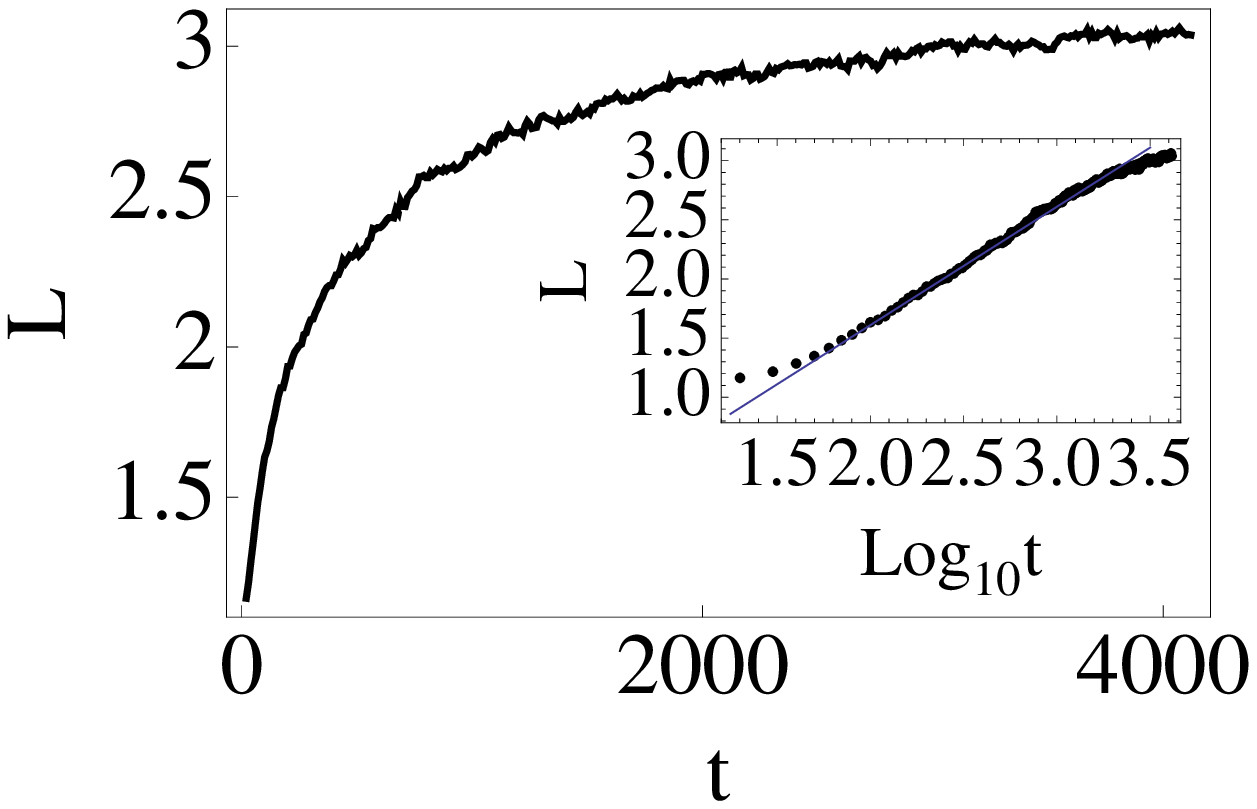}}
\put(100, -10){\includegraphics[scale=0.35]{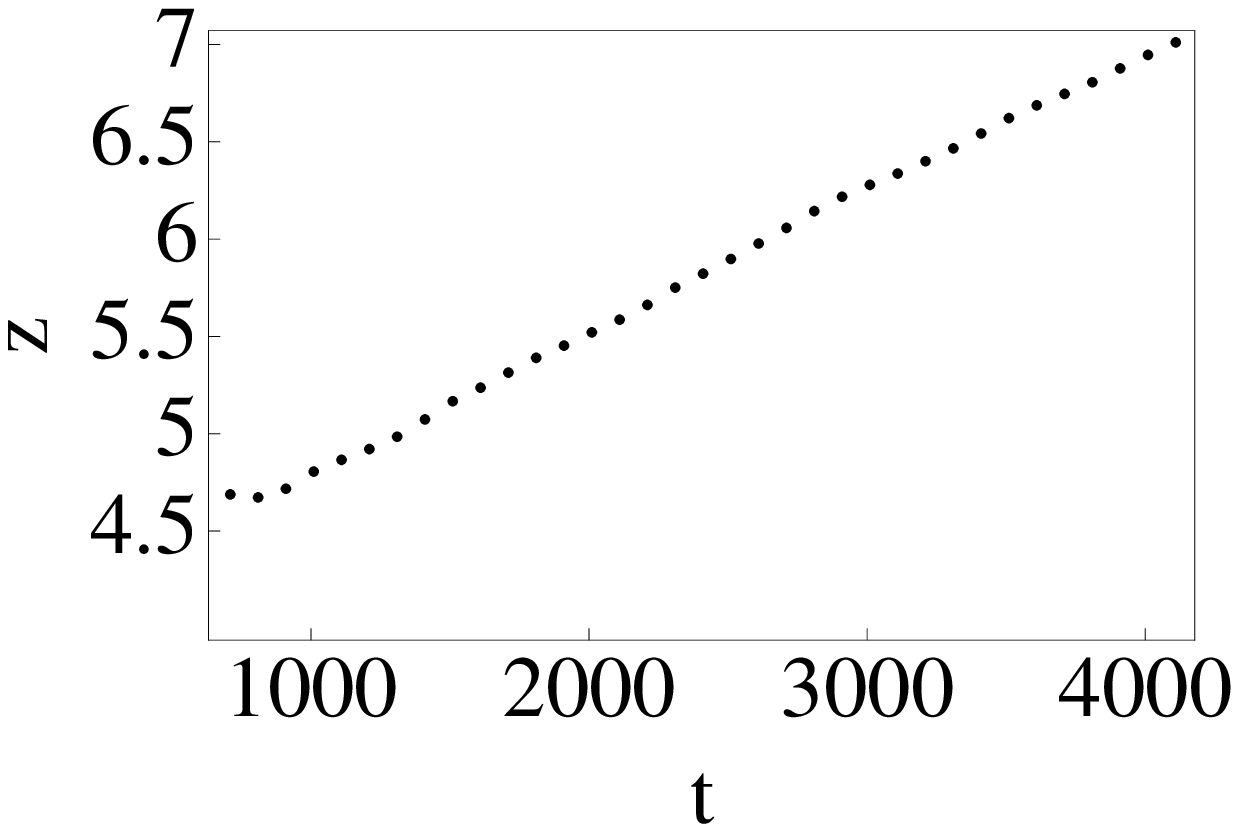}}
\end{picture}
\caption{\label{domanalysis} (Top-Left) Plot showing the evolution of the domain size ($L$) as a function of time. On short times the spin wave instability leads to the formation of local domains of size $L \sim 1$ lattice site. Over time these domains grow as shown in Fig.~\ref{coarsening}. (Inset) The data in the top figure is plotted against $\log_{10}(t)$. The solid line is a straight line fit to the data, indicating that the dynamics is consistent with a logarithmic growth law over two decades. (Top-Right) The data on the left is fit to the functional form $L(t) \sim t^{1/z}$ and the resulting $z$ is plotted as a function of time. The data indicates that the dynamics slows down over time.} \end{figure}

In Fig.~\ref{domanalysis}, we plot the average domain size as a function of time. Recall that for models with diffusion or advection dominated kinetics, the growth law takes on a power law form $L(t) \sim t^{1/z}$. For the curve shown in Fig.~\ref{domanalysis}, a fit to a power-law yields $z \sim 5$, much larger than the $z=3$ expected for diffusive dynamics of a conserved, scalar order parameter \cite{lifshitz}. Furthermore, as shown in the figure on the right, we find that  $z$ drifts to larger values over time, which suggests that the power law $L(t) \sim t^{1/3}$ is not recovered even in the infinite time limit \cite{huse}. 

The large value of $z$ points to a mechanism for transport other than advection or diffusion. Ao and Chui \cite{ao2} suggested that the late stages of spinodal decomposition in this system is predominantly driven by the Josephson effect, whereby domains grow when particles of one species tunnel across domains of the other species. As the domains grow larger, particles have to tunnel across a larger distance, which slows down the growth of domains. Assuming a simple model where the system is divided into a number of nearly fully polarized domains, they argued that the time to tunnel across domains $t \propto e^{L/\Lambda}$ where $L$ is the typical distance between two domains of the same species, and $\Lambda$ is the characteristic length-scale over which the wave functions of the two species overlap in equilibrium. Inverting this equation, one obtains $L/\Lambda \propto ln(t)$. In reality, there is a considerable spread in the polarization across the sample (see Fig.~\ref{magmagcorr}), and one would expect $L \sim (ln(t))^{\gamma}$ \cite{evans}.

For the parameters chosen in this simulation, $\Lambda \approx 1$ lattice site \cite{ao2}. In Fig.~\ref{domanalysis}, we plot $L(t)$ as a function of $\log_{10}(t)$ finding excellent agreement with a  straight line fit for roughly two decades ($\gamma \approx 1$). The deviations from this form can be attributed to the fact that the domains are only partially polarized. 

We caution the reader that the values of $z \sim 5$ and $\gamma \approx 1$ reported here depend on the cutoff $m_{\text{crit}}$. Increasing the cutoff flattens the curve for $L(t)$, making it harder to reliably extract $z$ from the data. For values of $m_{\text{crit}} < 0.5$, we have found that the dynamics is qualitatively similar to that reported here. For $m_{\text{crit}} \rightarrow 1$, one has to simulate the dynamics on much longer times to perform a similar analysis. 

To conclude, we have explored the dynamics of domain formation and coarsening in a binary Bose condensate, out of equilibrium. The parameters of the simulation are chosen such that the zero temperature mean-field ground state of the system is completely phase separated. However, on the timescales we study, we find that the density-density correlation function only develops appreciable on-site correlations, while longer range correlations remain negligible. The density-density correlation function is thus not a good indicator on the question of whether coarsening occurs in this system.  

We believe however that the system is indeed coarsening over time as illustrated in Fig.~\ref{coarsening}. We introduce a polarization cutoff and partition the system into domains of $a$ and $b$ atoms depending on whether the mean value of $|m(r)|$ in a region exceeds this cutoff. We find that over time, these domains indeed grow in size. However the rate of growth is considerably slower than the Lifshitz-Slyzov law of $L(t) \sim t^{1/z}$, where $z =3$ \cite{lifshitz}. Moreover, by calculating the drift in the exponent $z$, we showed that the dynamics appears to be slowing down in time, as $z$ drifts to larger values \cite{huse}. Our data seems to be consistent with \textit{tunneling} induced dynamics proposed by Ao and Chui \cite{ao2}, which yields \textit{logarithmic} growth law, with a dynamical exponent that is effectively infinite. 

In reality, there is always a small thermal cloud present in the trap. The physics discussed here can be extended to finite temperature by incorporating interactions between the condensate and the thermal cloud using stochastic versions of the Gross-Pitaevskii equation \cite{blakie}. The nature of coarsening dynamics resulting from these theories, and the associated growth laws is an important question for further study, and may eventually facilitate the comparison of models of coarsening dynamics with real experiments. 

\textit{Acknowledgements.---} SN would like to thank Ian Spielman for introducing Ref.~\cite{spielmanbinary} to him and Brandon Anderson for several discussions regarding the details of the numerical simulations performed in Ref.~\cite{spielmanbinary}. It is also a pleasure to thank Anushya Chandran, Michael Kolodrubetz and Ryan Wilson for enlightening discussions.
This work is supported by JQI-NSF-PFC, AFOSR-MURI, and ARO-MURI.

\end{document}